# Accuracy of the coherent potential approximation for a one-dimensional array with a Gaussian distribution of fluctuations in the on-site potential

I. Avgin

Department of Electrical and Electronics Engineering, Ege University, Bornova 35100,

Izmir, Turkey

A. Boukahil

Physics Department, University of Wisconsin-Whitewater, Whitewater, WI 53190

D. L. Huber

Physics Department, University of Wisconsin-Madison, Madison, WI 53706

We investigate the accuracy of the coherent potential approximation (CPA) for a one-dimensional array with nearest-neighbor interactions and a Gaussian distribution of fluctuations in the on-site potential. The CPA values of the integrated density of states and the inverse localization length are compared with the results of mode-counting studies carried out on arrays of $10^7 - 10^8$ sites. Good agreement is obtained suggesting that the CPA may be exact for this model. We also consider the asymptotic behavior of the inverse localization length and show that it can be approximated by the reciprocal of the decay length of a state localized about a single, strongly perturbed site in an otherwise perfect lattice.

# I. Introduction

Since its introduction more than forty years ago[1,2], the coherent potential approximation, or CPA, as it is commonly known, has proven to be a successful method of determining the distribution and properties of one-electron states in disordered materials. Due to its success, efforts have been made to determine if there are models for which the CPA gives exact results. Up to this point, the CPA has proven to be exact only for the Lloyd model where there is a Cauchy (or Lorentzian) distribution of the fluctuations in the on-site potential.[3,4] Early applications of the CPA to one- dimensional arrays having a Gaussian distribution of fluctuations have shown that the approximation is unusually effective for this model as well.[5,6] Because of this effectiveness, we have undertaken a systematic study of the accuracy of the CPA for the one-dimensional, nearest-neighbor model with Gaussian disorder. As will be discussed below, we found good agreement between the CPA predictions for integrated density of states and the inverse localization length and the corresponding results obtained by mode-counting in arrays of $10^7 - 10^8$ sites. The major source for the discrepancy between CPA and mode-counting appears to be errors in the numerical solution of the CPA equation, leading us to conjecture that the CPA may be exact for this model as well. As a by-product of this study, we investigate the asymptotic behavior of the inverse localization length and show that it is the characteristic of the decay rate of a wave function localized at a strongly perturbed site in an otherwise perfect array.

## II. Analysis

The Hamiltonian for the one-dimensional array takes the form

$$H = \sum_n V_n |n><n| - \sum_n [|n><n+1| + |n><n-1|] \qquad (1)$$

where the on-site potential, $V_n$, has a Gaussian distribution with zero mean and variance $\sigma^2$. Note that the nearest-neighbor coupling is equal to 1, so that in the absence of disorder, the band of eigenstates, symmetric about 0, ranges from $-2$ to 2.

In the coherent potential approximation, as applied to the system characterized by Eq.(1), the Green's function, $G_0^{CPA}(E)$ is expressed as

$$G_0^{CPA}(E) = \pi^{-1} \int_0^\pi dk [E + 2\cos(k) - V_c(E)]^{-1} \qquad (2)$$

where the coherent potential, $V_c(E)$, satisfies the equation

$$\int dV P(V)[V - V_c(E)] / [1 - (V - V_c(E)) G_0^{CPA}(E)] = 0 \qquad (3)$$

with $P(V)$ having the Gaussian form.

The main focus of this paper is on assessing the accuracy of the coherent potential approximation for the single-particle Green's function by comparing the integral of $G_0^{CPA}(E)$ with accurate numerical results obtained from large arrays. A straightforward way to do this would be to diagonalize the Hamiltonian and calculate the Green's function directly from the eigenvalues. This approach is limited, however, by number of sites in the array (typically, $10^3 - 10^4$). In one dimensional arrays with nearest-neighbor interactions, one can make use of mode-counting techniques to establish accuracy of the CPA in arrays of $10^8$ (or more) sites.

Information about the real and imaginary parts of the Green's function can be obtained from an analysis of the sequence of amplitude ratios, $R_n = a_n/a_{n-1}$, associated with the eigenstates of the Hamiltonian[7]

$$R_n(E) = E - V_n - 1/R_{n-1}(E) \tag{4}$$

with $N$ being the number of sites, the number of sign changes in the sequence $R_1(E)$ ($= E - V_1$), $R_2(E)$, ..., $R_N(E)$ corresponds to the number of modes with energies less than $E$. When $N \gg 1$, which is the case here, we identify the number of sign changes, divided by $N$, with the integrated density of states per site, $IDOS(E)$. The integrated density of states can also be obtained from the imaginary part of the Green's function using the equation

$$IDOS(E) - IDOS(0) = \pi^{-1} \int_0^E dE' \, \text{Im} \, G_0(E') \tag{5}$$

Information about the spatial extent of the wave functions follows from a consideration of the logarithm of $|R_n(E)|$.[8,9] The inverse localization length, $ILL(E)$, which is identified with the reciprocal of the average fall-off distance for eigenstates with energy $E$, is expressed as[8]

$$ILL(E) = -N^{-1} < \ln|a_N/a_1|>_E$$

$$= -N^{-1} \ln \left| \prod_{n=1}^{N} R_n(E) \right| = -N^{-1} \sum_{n=1}^{N} \ln|R_n(E)| \quad N \to \infty \tag{6}$$

The equivalent expression for $ILL(E)$ involving the Green's function utilizes the real part of $G_0(E)$. It can be derived by integrating the real part of the spectral representation for $G_0(E)$ and making use of the expression for the inverse localization length derived in Ref. 8

$$ILL(E) = \int dx \rho(x) \ln|E - x| \qquad (7)$$

in which $\rho(x)$ denotes the density of states. The equation takes the form

$$ILL(E) - ILL(0) = \int_0^E dE' \, \text{Re}(G_0(E')) \qquad (8)$$

In assessing the accuracy of the CPA, we focus on $IDOS(E) - IDOS(0)$ and $ILL(E) - ILL(0)$ whose derivatives with respect to $E$ yield the real and imaginary parts of $G_0(E)$.

In Figs. 1 and 2, we compare the mode-counting and CPA results for the integrals of the imaginary (Fig. 1 and Eq. (5)) and real (Fig. 2 and Eq. (8)) parts of the Green's functions calculated with $\sigma^2 = 0.25$, 1.0, and 4.0. The numerical results for the *IDOS* involved counting the number of sign changes in the sequence, $R_1, R_2, \ldots, R_N$, divided by $N$, whereas the values for the *ILL* were obtained from Eq. (6). In both cases, $N = 4 \times 10^7$. The values of $IDOS(0)$ and $ILL(0)$ were 0.50002 and 0.02821 for $\sigma^2 = 0.25$, 0.49999 and 0.10880 for $\sigma^2 = 1.0$, and 0.50012 and 0.35731 for $\sigma^2 = 4.0$. The CPA results were obtained from by solving Eqs. (2) and (3) for $G_0^{CPA}(E)$ which is then used in the evaluation of the right hand side of Eqs. (5) and (8). Standard MatLab programs in double precision were used in the solution of the self-consistent equation for the coherent potential and in the evaluation of the integrals.

We note that the $IDOS(E) - IDOS(0)$ curves approach 1/2 as $E \to \infty$, consistent with the fact that the number of modes is equal to the number of sites. In the limit $N \to \infty$, $IDOS(0) = \frac{1}{2}$ since there are equal numbers of positive and negative energy modes. The asymptotic behavior of the *ILL* is more interesting. In the limit of large $|E|$, *ILL* approaches the reciprocal of the decay length of a localized state with energy $E$ associated with a perturbed site in an otherwise unperturbed lattice. The connection between the energy of the localized state and the shift in the on-site potential, $V_0$, is through the equation[10, 11]

$$E = \pm 2(1 + (V_0/2)^2)^{1/2} \qquad (9)$$

where the sign corresponds to the sign of $V_0$. The reciprocal of the decay length, $\delta(E)$, (in units of the inverse lattice constant) associated with the localized state is given by

$$1/\delta(E) = -\ln\{(|E/2|)[1-(1-4/E^2)^{1/2}]\} \qquad (10)$$

which is appropriate only for $|E| > 2$ and has the limiting behavior $1/\delta(E) \to \ln|E|$ for large $|E|$. In Fig. 3 we compare the mode-counting values of the $ILL(E)$ with the results obtained by approximating $ILL(E)$ as $1/\delta(E)$, which we refer to as the 'single-site approximation'. In contrast to $IDOS(0)$, there appears to be no simple analytical expression for $ILL(0)$. We postpone consideration of these results until the following section.

## III. Discussion

It is evident from Figs. 1 and 2 that there is good agreement between the numerical results and the results obtained using the coherent potential approximation. The extent of the agreement for $\sigma^2 = 1.0$ is shown in greater detail in Fig. 4 where we plot the relative difference

$$\Delta IDOS = [IDOS(E)_{\text{mode-counting}} - IDOS(E)_{\text{CPA}}]/IDOS(E)_{\text{mode-counting}}$$

and in Fig. 5 where we plot

$$\Delta ILL = [ILL(E)_{\text{mode-counting}} - ILL(E)_{\text{CPA}}]/ILL(E)_{\text{mode-counting}}$$

where in both cases $N = 10^8$. In Figs. 4 and 5 it is apparent that $|\Delta IDOS(E)| < 2.5 \times 10^{-3}$ whereas $|\Delta ILL(E)| < 5.2 \times 10^{-2}$. Since we have carried out similar computations with a different sequence of potential fluctuations and obtained results that were indistinguishable from those shown, we suspect that the qualitative difference between the *IDOS* and the *ILL* is connected with numerical solution of the self-consistent equation for the coherent potential.

The results displayed in Fig. 3, along with Eq. (9), indicate that the single-site approximation works well when the energy of the localized state lies in a region where the probability of the corresponding potential fluctuation $V_0 = \pm 2[(E/2)^2 - 1]$, which is expressed as

$$P(V_0) = (2\pi\sigma^2)^{-1/2} \exp\{-2[(E/2)^2 - 1]/\sigma^2\} \approx (2\pi\sigma^2)^{-1/2} \exp[-E^2/2\sigma^2] \quad (11)$$

is extremely small, i.e. $P(V_0)/P(0) \approx \exp[-E^2/2\sigma^2] \ll 1$. It should be noted that the single-site approximation is not limited to the Gaussian distribution, but is applicable to other distributions in the large-$E$ limit, e.g. the Cauchy distribution, as can be seen from the exact expression for the *ILL* of the Cauchy distribution given in Ref. 8.

Taken together, the results presented in this paper are evidence that the CPA is a good approximate theory for the single-particle Green's function for the one-dimensional Gaussian model and lead us to conjecture that it may be exact. To strengthen the argument, it is necessary to make significant improvements in the accuracy of the numerical solutions of the CPA equations, although only analytical studies can establish that the coherent potential approximation is exact.

**Acknowledgment**

I. Avgin would like to thank the Scientific and Technological Research Council of Turkey (TUBITAK) for their partial support.

**Figure Captions**

Fig. 1  $IDOS(E) - IDOS(0)$ vs $E$.  The solid curves are data obtained by mode-counting for an array of $4 \times 10^7$ sites.  The symbols denote the values obtained from the CPA.  $\sigma = 0.5, 1.0,$ and $2.0$.

Fig. 2  $ILL(E) - ILL(0)$ vs $E$.  The solid curves are data obtained by mode-counting for an array of $4 \times 10^7$ sites.  The symbols denote the values obtained from the CPA.  $\sigma = 0.5, 1.0,$ and $2.0$.

Fig. 3  $ILL(E)$ vs $E$.  Comparison between mode-counting results obtained from an array of $4 \times 10^7$ sites, shown as data points for $\sigma = 0.5, 1.0,$ and $2.0$, and the single-site approximation, $ILL(E) = 1/\delta(E)$, shown as a solid curve.

Fig. 4  $\Delta IDOS(E) = [IDOS(E)_{\text{mode-counting}} - IDOS(E)_{\text{CPA}}]/IDOS(E)_{\text{mode-counting}}$ vs $E$ for $\sigma = 1.0$.  Array size $= 10^8$ sites.

Fig. 5  $\Delta ILL(E) = [ILL(E)_{\text{mode-counting}} - ILL(E)_{\text{CPA}}]/ILL(E)_{\text{mode-counting}}$ vs $E$ for $\sigma = 1.0$.  Array size $= 10^8$ sites.

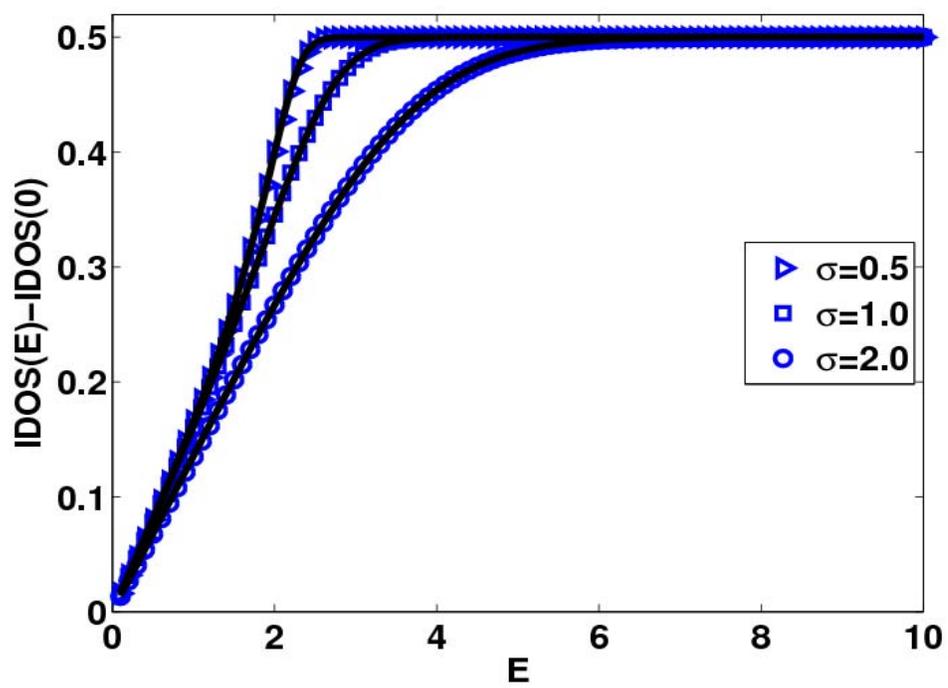

Fig. 1

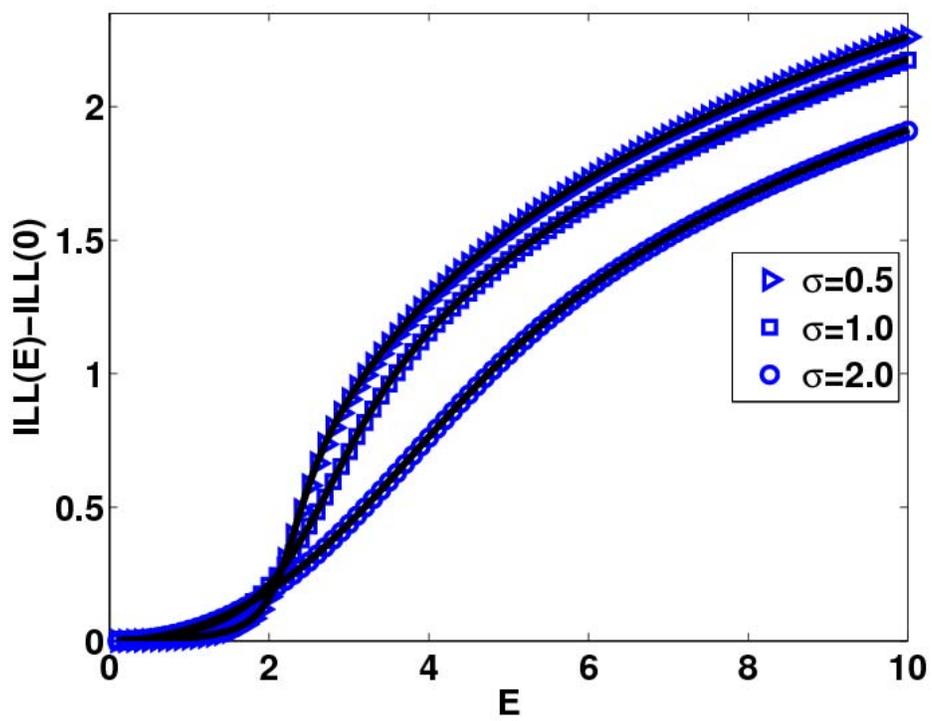

Fig. 2

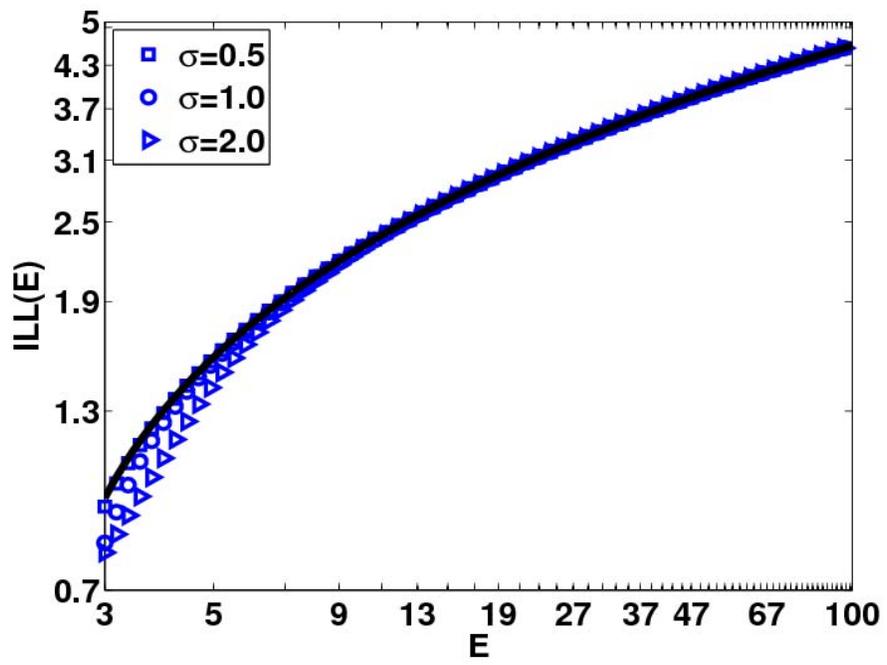

Fig. 3

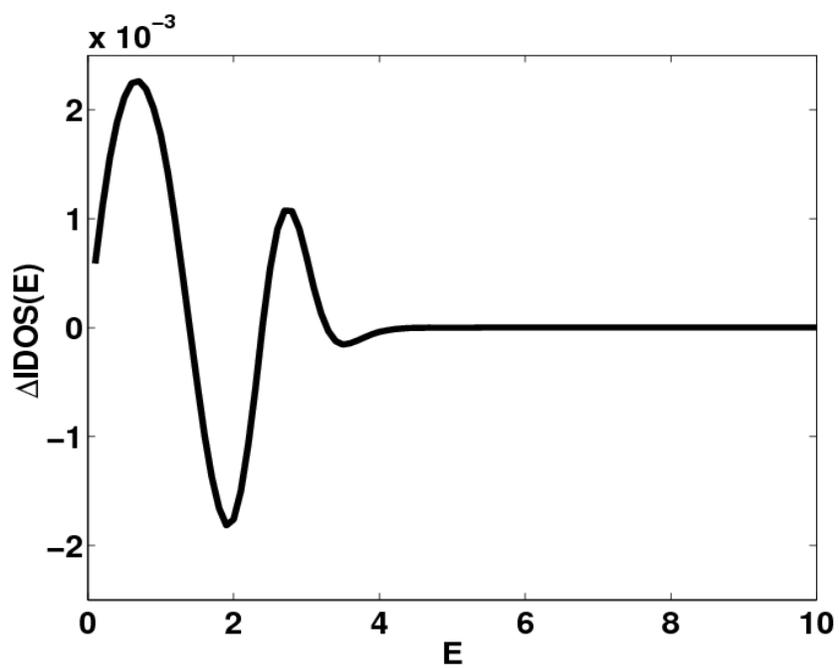

Fig. 4

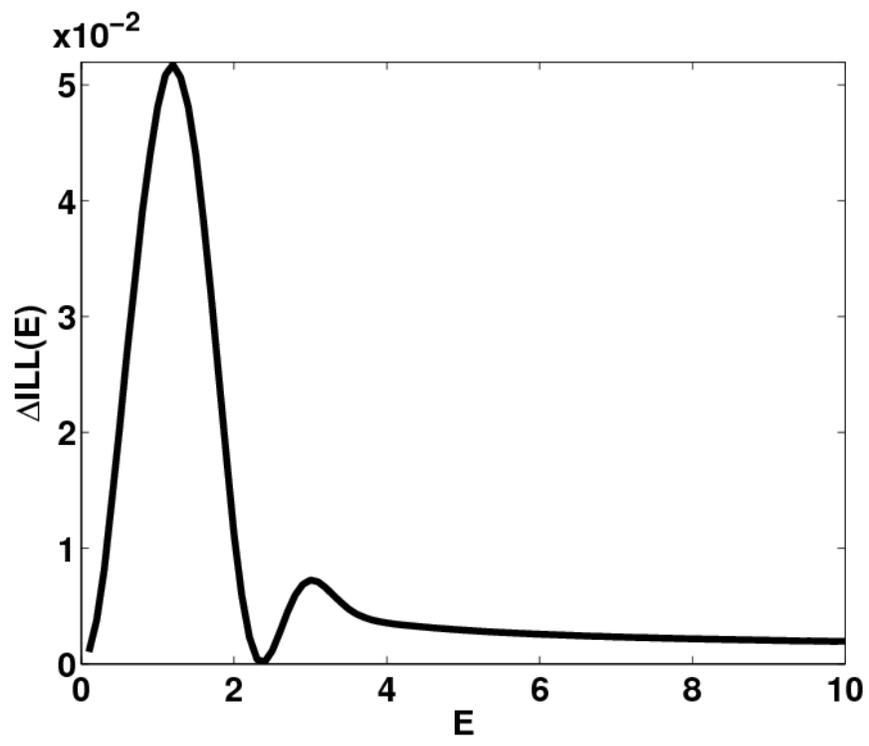

Fig. 5